\documentclass[aps,prd,showpacs,nofootinbib,superscriptaddress]{revtex4}
\usepackage{bm,graphicx,amssymb}
\begin{document}
\title{Gauge invariance and form factors for the decay $\bm{ B\to \gamma l^+l^-}$}
\author{Frank Kr\"uger}
\email{fkrueger@ph.tum.de}
\affiliation{Physik Department, Technische Universit\"at M\"unchen, 
D-85748 Garching, Germany}
\author{Dmitri Melikhov}
\email{melikhov@thphys.uni-heidelberg.de}
\affiliation{Institut f\"ur Theoretische Physik, Universit\"at Heidelberg,  
Philosophenweg 16,  D-69120, Heidelberg, Germany}
\begin{abstract}
We analyse the form factors for the $B\to \gamma l^+l^-$ weak transition. We show that 
making use of the gauge invariance of the $B\to\gamma l^+l^-$ amplitude, the structure 
of the form factors in the resonance region, and their relations at large values of the 
photon energy results in efficient constraints on the behavior of the form factors.  
Based on these constraints, we propose a simple parametrization of the form factors 
and apply it to the lepton forward-backward (FB) asymmetry in the $B\to\gamma l^+l^-$ decay. 
We find that the behavior of the FB asymmetry 
as a function of the photon energy, 
as well as the location of its zero, depend only weakly on the 
$B\to\gamma$ form factors, 
and thus constitutes a powerful tool for testing 
the standard model. 
\end{abstract}
\maketitle

\section{Introduction}
Recently, the decay $B\to\gamma l^+l^-$ has been the subject of 
a number of investigations \cite{sehgal,bllg:decay,asym,aliev,geng}, where it has been pointed out 
that this process may serve as an important probe of the standard model (SM) and 
possible extensions. However, knowledge of the long-distance QCD effects, 
which are inherently non-perturbative, is important 
to extract quantitative information on the underlying short-distance interactions. 

From the analyses of the decay $B\to K^* l^+l^-$ it is known \cite{silvano,gustavo}
that the uncertainties due to the hadronic  form factors are considerably reduced 
if one considers asymmetries such as the forward-backward (FB) 
asymmetry of the lepton.
This empirical observation later received an explanation within the 
large energy effective theory (LEET) \cite{leet}. According to LEET, all the 
heavy-to-light meson transition form factors are given 
at leading order in $1/M_B$ and $1/E$ ($E$ is the energy of the light meson)
in terms of a few universal form factors, and thanks to that the form factor effects
largely drop out from the asymmetries. 

As for the radiative dilepton decay $B\to\gamma l^+l^-$, one expects the same to 
be true. However, a surprisingly strong 
dependence of the FB asymmetry on the specific form factor model can be found in the 
existing literature \cite{sehgal,bllg:decay,asym}, which needs better understanding.  

In this paper, we analyse the form factors for the $B\to\gamma$ transition induced 
by vector, axial-vector, tensor, and pseudotensor currents. 

We  show that important relations between form 
factors of different currents arise as a consequence of the gauge invariance of the 
$B\to\gamma$ amplitude. We derive an exact relation between the 
form factors of tensor and pseudotensor currents at $q^2=0$, where 
$q^2$ is the dilepton invariant mass in the decay $B\to \gamma l^+l^-$. 
We note that the form factors from a recent sum-rule calculation of Ref. \cite{aliev} 
are inconsistent with this exact relation.  
 
We investigate the behavior of the various form factors at large $q^2$ and find 
interesting relations corresponding to the resonance contributions to the form factors. 
We  argue that these contributions signal substantial corrections 
to the Isgur-Wise relations valid to $1/m_b$ accuracy at large $q^2$ \cite{iw}. 
 
Combined with the relations among the form factors from LEET, the results obtained 
provide 
strong restrictions on the $B\to\gamma$ form factors. We propose a simple 
model for the form factors which is valid over the full range of the photon 
energy, and which satisfies all known constraints. 
An important remark is in order here: It has been shown recently that a proper account 
of collinear and soft gluons leads to a different effective theory -- the so-called 
soft-collinear effective theory (SCET) \cite{cset1} which supersedes the LEET. Important for our discussion is 
that interactions with collinear gluons preserve the {\it relations} for the soft part of 
the heavy-to-light form factors from LEET \cite{cset2} 
[the differences appear in the $O(\alpha_s)$ part]. Our analysis is therefore fully compatible with SCET.

As an application of our form factor model, 
we examine the FB asymmetry in the decay ${B}_s\to\gamma \mu^+\mu^-$. 
We find that this asymmetry, and particularly its zero arising in the SM, 
can be predicted with small theoretical uncertainties. 

\section{\label{form:factors}Form factors for the $\bm{B\to \gamma}$ transition}
We are concerned with  the amplitudes of the $B\to\gamma$ transition 
resulting from various quark currents. In our analysis, we adopt the following 
conventions: 
\begin{eqnarray}
\gamma^{5}=i\gamma^{0}\gamma^{1}\gamma^{2}\gamma^{3},
\qquad \sigma_{\mu \nu}={\frac{i}{2}}[\gamma_{\mu},\gamma_{\nu}], \qquad
\epsilon^{0123}=-1, 
\end{eqnarray}
and accordingly 
\begin{eqnarray}
\label{2}
\sigma_{\mu\nu}\gamma^5=-\frac{i}{2}\epsilon_{\mu\nu\alpha\beta}
\sigma^{\alpha\beta}.
\end{eqnarray}

\subsection{Transition to a virtual photon} 
We start with the amplitude describing the transition of the 
$B_q$  ($q=s,d$) 
meson with momentum $p$ 
to a virtual photon with momentum $k$.
In this case, the form factors depend 
on two variables: that is, the photon virtuality $k^2$ and the square of the 
momentum transfer $(p-k)^2$. 
As we shall see, gauge invariance and the absence of singularities in the amplitude 
lead to several relations among the form factors at $k^2=0$, thereby  reducing the 
number of independent form factors for the transition to a  \emph{real} photon.  


(i) For the $B_q\to \gamma^*$ transition induced by the \emph{axial-vector current}, 
the gauge-invariant amplitude (with respect to the photon) contains three form factors 
and can be written in the form\footnote{
Notice that for the 
$B$-meson transition to a
virtual photon  the amplitude of the axial-vector current contains a contact term, which is proportional to the charge of the $B$ meson. The contact term is thus 
present for the {\it charged} $B$-meson transition, but is absent in the case of a 
\emph{neutral} $B$ meson. For a detailed discussion of form factors and contact 
terms in this amplitude, see Ref.~\cite{ms}.}  
\begin{eqnarray}
\label{axial-vector}
\langle \gamma^*(k)|\bar q \gamma_\mu\gamma_5 b|\bar B_q(p)\rangle
= ie\varepsilon^{\ast\alpha}(k)
\left\{
\left(
g_{\mu\alpha}-\frac{k_\alpha k_{\mu}}{k^2}\right)f
+
p_{\mu}\left(p_{\alpha}-\frac{k\cdot p}{k^2}k_\alpha\right)a_1
+ 
k_{\mu}\left(p_{\alpha}-\frac{k\cdot p}{k^2}k_\alpha\right)a_2 
\right\},
\end{eqnarray} 
where  we have explicitly written the gauge-invariant 
Lorentz structures. In the above, $\varepsilon^\alpha$ denotes the 
polarization vector of the photon, $e=\sqrt{4\pi\alpha}$, and the form factors are 
defined according to Ref.~\cite{ms}. Since the amplitude is a regular function at 
$k^2=0$, the requirement of gauge invariance results in 
the following constraints on the form factors at $k^2=0$:  
\begin{eqnarray}
\label{constraint1}
f+(k\cdot p)\ a_2=0,\qquad a_1=0. 
\end{eqnarray} 


(ii) For the transition induced by the \emph{vector current}, the amplitude is
para\-me\-trized in terms of a
single form factor $g$; namely, 
\begin{eqnarray}
\label{vector}
\langle \gamma^*(k)|\bar q\gamma_\mu b|\bar B_q(p)\rangle &=& 
2e g \varepsilon^{\ast\alpha}(k)
\epsilon_{\mu\alpha\rho\sigma} p^{\rho} k^{\sigma}.
\end{eqnarray}


(iii) For the $B_q\to \gamma^*$ transition 
induced by the \emph{pseudotensor current}, the amplitude can be written 
in terms of three form factors:
\begin{eqnarray}
\label{pseudotensor}
\langle \gamma^*(k)|\bar q \sigma_{\mu\nu}\gamma_5 b|\bar B_q(p) \rangle
&=&e\varepsilon^{*\alpha}(k)
\bigg\{
\bigg[
\bigg(g_{\alpha\nu}-\frac{k_\alpha k_{\nu}}{k^2}\bigg)p_\mu-
\bigg(g_{\alpha\mu}-\frac{k_\alpha k_{\mu}}{k^2}\bigg)p_\nu\bigg]g_1
\nonumber\\
&&\hspace{4.em}\mbox{}
+
(g_{\alpha\nu}k_\mu-g_{\alpha\mu}k_\nu)g_2
+
\bigg(p_{\alpha}-\frac{k\cdot p}{k^2}k_\alpha\bigg)(k_\mu p_\nu-p_\nu k_\mu)g_0
\bigg\}.
\end{eqnarray}
At $k^2=0$, gauge invariance leads to the condition 
\begin{eqnarray}
\label{constraint2}
g_1-(k\cdot p)\ g_0=0. 
\end{eqnarray} 


(iv) The amplitude for the transition induced by the \emph{tensor current} 
can be  obtained 
from Eq.~(\ref{pseudotensor}) by applying the identity in Eq.~(\ref{2}),  
and is given by
\begin{eqnarray}
\label{tensor}
\langle \gamma^*(k)|\bar q \sigma_{\mu\nu} b|\bar B_q(p)\rangle 
&=& ie\varepsilon^{*\alpha}(k)
\bigg\{
\bigg(\epsilon_{\mu\nu\alpha \rho}p^\rho 
-\frac{k_\alpha}{k^2}\epsilon_{\mu\nu \sigma\rho}k^\sigma p^\rho\bigg)
g_1+
\epsilon_{\mu\nu\alpha \sigma}k^\sigma g_2
+\bigg(p_{\alpha}-\frac{k\cdot p}{k^2}k_\alpha\bigg)\epsilon_{\mu\nu \rho\sigma}p^\rho k^\sigma g_0
\bigg\}. 
\end{eqnarray} 

\subsection{Transition to a real photon} 
For the transition to a real photon, the matrix 
element of the vector current is given by Eq.~(\ref{vector}) while 
the amplitude for the axial-vector current, employing 
the relation in Eq.~(\ref{constraint1}), reads
\begin{eqnarray}
\label{axial-vector1}
\langle \gamma(k)|\bar q \gamma_\mu\gamma_5 b|\bar B_q(p)\rangle
=-ie\varepsilon^{\ast\alpha}(k)
[
g_{\mu\alpha}(p\cdot k)-p_\alpha k_{\mu}]a_2 (k^2=0). 
\end{eqnarray} 
As for tensor and pseudotensor currents, their matrix elements have the form 
\begin{eqnarray}
\label{tensor1}
\langle \gamma (k)|\bar q \sigma_{\mu\nu}\gamma_5 b|\bar B_q(p) \rangle
&=&e\varepsilon^{*\alpha}(k)
\bigg\{
\bigg(g_{\alpha\nu}k_\mu-g_{\alpha\mu}k_\nu\bigg)g_2
+
\bigg(
[g_{\alpha\nu}(p\cdot k)-p_\alpha k_{\nu}]p_\mu-
[g_{\alpha\mu}(p\cdot k)-p_\alpha k_{\mu}]p_\nu\bigg)g_0 \bigg\},
\nonumber
\\
\langle \gamma (k)|\bar q \sigma_{\mu\nu} b|\bar B_q(p)\rangle 
&=&ie\varepsilon^{*\alpha}(k)
\bigg\{
\epsilon_{\mu\nu\alpha \sigma}k^\sigma g_2
+\bigg[p_{\alpha}\epsilon_{\mu\nu \rho\sigma}p^\rho k^\sigma 
-(p\cdot k)\epsilon_{\mu\nu \rho\alpha}p^\rho\bigg]g_0
\bigg\}. 
\end{eqnarray}  
It should be noted that in the case of a real photon the 
gauge-invariant amplitudes of the pseudotensor and 
tensor currents contain two Lorentz structures and not only one as stated 
in Ref.~\cite{korch}.  
Multiplying Eq.~(\ref{tensor1}) by $(p-k)^\nu$, we arrive at the following set of 
amplitudes that describe the $B\to\gamma$ transition:  
\begin{eqnarray}
\label{real}
\langle \gamma(k)|\bar q \gamma_\mu\gamma_5 b|\bar B_q(p) \rangle&=&
ie\varepsilon^{\ast\alpha}(k)
[g_{\mu\alpha} (p\cdot k)-p_\alpha k_\mu]\frac{F_A}{M_{B_q}}, 
\nonumber
\\
\langle \gamma(k)|\bar q\gamma_\mu b|\bar B_q(p)\rangle &=& 
e\varepsilon^{\ast\alpha}(k)\epsilon_{\mu\alpha\rho\sigma} p^{\rho} k^{\sigma}\frac{F_V}{M_{B_q}},   
\nonumber
\\
\langle \gamma(k)|\bar q \sigma_{\mu\nu}\gamma_5 b|\bar B_q(p) \rangle (p-k)^\nu
&=& 
e\varepsilon^{\ast \alpha}(k)[g_{\mu\alpha}(p\cdot k)- p_{\alpha}k_{\mu}]F_{TA}, 
\nonumber
\\
\langle \gamma(k)|\bar q \sigma_{\mu\nu} b|\bar B_q(p) \rangle (p-k)^\nu
&=& 
ie\varepsilon^{\ast \alpha}(k)\epsilon_{\mu\alpha\rho\sigma}p^\rho k^\sigma F_{TV}, 
\end{eqnarray}
where the dimensionless form factors $F_i$ are given in terms of the form factors 
$g$, $a_2$, $g_0$, $g_2$ at $k^2=0$:
\begin{eqnarray}
\label{rel1}
&&F_V=2M_{B_q} g,\qquad  F_A=-M_{B_q} a_2, \\ \label{rel2}
&&F_{TV}=-g_2-[p^2-(p\cdot k)] g_0,\qquad  F_{TA}=-g_2-(p\cdot k) g_0.
\end{eqnarray}

For a real photon in the final state, the form factors depend on the square of 
the momentum transfer, $q^2\equiv (p-k)^2$. 
Equivalently, one may consider the form factors 
as functions of the photon energy in the $B$-meson rest frame, 
\begin{eqnarray}
E=\frac{M_B}{2} \left(1-\frac{q^2}{M_B^2}\right). 
\end{eqnarray}
For massless leptons, the kinematically accessible range is 
\begin{eqnarray}
0\le  q^2\le M_B^2,\qquad 0\le E\le E_{\mathrm{max}}=M_B/2. 
\end{eqnarray}
Then, rewriting Eq.~(\ref{rel2}) in the form 
\begin{eqnarray}
\label{exact1}
F_{TV}=F_{TA}-M_B(M_B-2 E)g_0,
\end{eqnarray}
we derive the following \emph{exact} relation:  
\begin{eqnarray}
\label{exact2}
F_{TA}(E_{\mathrm{max}})&=&F_{TV}(E_{\mathrm{max}}). 
\end{eqnarray}
We stress that the equality of the 
form factors $F_{TA}$ and $F_{TV}$ is  valid only at 
 $E=E_{\mathrm{max}}$.

\subsection{LEET and form factors at large $\bm{E}$}
Interesting relations between the form factors emerge in the limit 
where the initial hadron is heavy and the final photon has a large energy \cite{leet}.  
In this case, the form factors may be expanded in  inverse powers of $\Lambda_{\mathrm{QCD}}/M_B$ 
and $\Lambda_{\mathrm{QCD}}/E$. As a result, to leading-order accuracy, one finds  
\begin{eqnarray}
\label{leet}
F_V\simeq F_A \simeq F_{TA}\simeq F_{TV}\simeq \zeta^\gamma_\perp(E,M_B), 
\end{eqnarray}
where $\zeta^\gamma_\perp(E,M_B)$ is the universal form factor for the $B_q\to\gamma$ 
transition (cf.~Ref.~\cite{leet}).\footnote{
For a massive particle in the final state one has  
two universal form factors 
$\zeta_\perp(E,M_B)$ and $\zeta_{||}(E,M_B)$. The latter does not contribute 
in the case of a massless final vector particle. 
We also note the relation 
$\zeta^{B_u\to\gamma}_\perp(E,M_B)=Q_u/Q_d\;\zeta^{B_d\to\gamma}_\perp(E,M_B)$, 
where $Q_u/Q_d=-2$.} 
We emphasize that this relation is violated by terms of order  
$O(\Lambda_{\mathrm{QCD}}/M_B)$ and  
$O(\Lambda_{\mathrm{QCD}}/E)$, as well as by radiative corrections of  
$O(\alpha_s)$ \cite{beneke}. 

For large values of $E$, the energy dependence of the universal $B\to\gamma$ form factor 
can be obtained from perturbative QCD, the result being \cite{korch}
\begin{eqnarray}
\label{korch}
\zeta^\gamma_\perp(E,M_B)\propto  {f_B M_B}/{E}. 
\end{eqnarray}

\subsection{Heavy quark symmetry and form factors at large $\bm{q^2}$}
As found by Isgur and Wise \cite{iw}, in the region of large $q^2\simeq M_B^2$ the 
form factors satisfy the relations
\begin{eqnarray}
\label{iw}
g_2\simeq -2M_B g,\qquad g_0\simeq \frac{1}{2M_B}(2g+a_2),
\end{eqnarray}
which are valid at leading order in $\Lambda_{\rm QCD}/m_b$.\footnote{
It should be noted that at large $q^2$ there are in general three 
independent relations between the 
$B\to V$ form factors. 
The third relation, however,  is automatically satisfied in the case of a photon in the 
final state, due to the gauge-invariance constraints in 
Eqs.~(\ref{constraint1}) and (\ref{constraint2}).}  
Combining these expressions with Eqs.~(\ref{rel1}) and (\ref{rel2}), we derive  
the following result for the tensor-type form factors in the region of small $E$:
\begin{eqnarray}
\label{FF}
\nonumber
F_{TV}&\simeq &F_V-\frac{M_B-E}{2M_B}(F_V-F_A),\\
F_{TA}&\simeq &F_V-\frac{E}{2M_B}(F_V-F_A). 
\end{eqnarray}
It is obvious that these relations are compatible with those in 
Eq.~(\ref{leet}), which emerge in the region where $E$ is large, and hence 
are valid to leading order in 
$\Lambda_{\rm QCD}/m_b$ in the full range of $E$.
Yet the leading-order $\Lambda_{\rm QCD}/m_b$ relations in Eq.~(\ref{FF}) 
may not be sufficient to understand the behavior of the form factors $F_{V,A}$ and 
$F_{TA,TV}$ in the region where $q^2\simeq M_B^2$. 

To explain this point, 
it is useful to express the form factors $F_i$ in terms of the Wirbel-Stech-Bauer 
(WSB) form factors \cite{wsb}. The advantage of the WSB form factors is that 
each one has definite spin and parity, and hence contains 
contributions of resonances with the corresponding quantum numbers. Explicitly, we 
have 
\begin{eqnarray}
\label{wsb}
F_V=2V^{B\to\gamma}, \qquad 
F_A=\frac{M_B}{E}A_1^{B\to\gamma}, \qquad 
F_{TV}=2T_1^{B\to\gamma}, \qquad 
F_{TA}=\frac{M_B}{E}T_2^{B\to\gamma}. 
\end{eqnarray}
Notice that the relation $F_{TV}=F_{TA}$ at $q^2=0$ (or $E=M_B/2$) is 
just the well-known relation $T_1(0)=T_2(0)$. 
Then, making use of the  constraints in 
Eqs.~(\ref{constraint1}) and (\ref{constraint2}), which are due to 
electromagnetic gauge invariance, we obtain exact relations between 
the WSB form factors that are relevant for the transition into a real photon; namely,
\begin{eqnarray}
\label{wsb1}
T_2^{B\to\gamma}=\frac{2E}{M_B}T_3^{B\to\gamma}, 
\qquad   
A_1^{B\to\gamma}=\frac{2E}{M_B}A_2^{B\to\gamma}.
\end{eqnarray}
By virtue of these relations, we may rewrite Eq.~(\ref{wsb}) in the form 
\begin{eqnarray}
\label{wsb2}
F_A=2A_2^{B\to\gamma},\qquad F_{TA}=2T_3^{B\to\gamma}, 
\end{eqnarray}
which exhibits the absence of a singular behavior of $F_A$ and $F_{TA}$. 

We now examine the analytic structure of the form factors near $q^2=M_B^2$, starting
with $V$ and $T_1$, which have a pole at $q^2=M^2_{B^*}$.
This pole is located very close to 
the upper boundary of the physical region, $q^2=M_B^2$, since $M_{B^*}-M_B=45\;{\rm MeV}\sim O(\Lambda_{\mathrm{QCD}}^2/m_b)$. 
Moreover, as shown in Refs.~\cite{ms,stech}, the residues of the form factors $V$ and $T_1$ 
in the pole at $q^2=M^2_{B^*}$ are equal in the heavy quark limit $m_b\to\infty$. 
As a consequence, $F_V$ and $F_{TV}$ should be approximately equal 
and rise steeply 
as $q^2\to M_B^2$.  

As for the form factors $F_A$ and $F_{TA}$, they are expected to have 
qualitatively different behavior 
near $q^2=M_B^2$. Indeed, the masses of the resonances that correspond to $A_2$ and 
$T_3$ [Eq.~(\ref{wsb2})], denoted by $B^{**}$, are expected to be several hundred MeV higher than $M_B$, since
$M_{B^{**}}-M_B\sim O(\Lambda_{\mathrm{QCD}})$. Thus, 
singularities of the form factors $F_A$ and $F_{TA}$ 
are much farther 
from the physical region, compared to the form factors $F_{V}$ and $F_{TV}$. 
Consequently, $F_A$ and $F_{TA}$ are relatively flat as $q^2\to M_B^2$. 

It is now clear why the leading-order $\Lambda_{\mathrm{QCD}}/m_b$ 
relations (\ref{FF}) are not useful for  
understanding the behavior of the form factors near $q^2=M_B^2$. As a matter of fact, 
at leading order in 
$\Lambda_{\mathrm{QCD}}/m_b$ all the $b\bar q$ resonances with different spins have 
the same masses, so
that all the form factors $F_{i}$ have poles at $q^2=M_B^2$ (i.e., at $E=0$). 
This picture is fully consistent with Eq.~(\ref{FF}), but it is far from reality,   
since $O(\Lambda_{\mathrm{QCD}}/m_b)$ corrections to the form factor relations in 
Eq.~(\ref{FF}) become crucial in the region near $q^2=M_B^2$. 
We expect, then, the following relation between the form factors near $q^2=M_B^2$:
\begin{eqnarray}
F_A\simeq F_{TA} \ll F_V\simeq F_{TV},
\end{eqnarray}
in agreement with the resonance location.

\subsection{The general picture of the $\bm{B\to\gamma}$ form factors}
Combining the above information on the form factors, 
the following picture of the $B\to\gamma$ form factors emerges.

\begin{itemize}
\item At $E=E_{\mathrm{max}}$, the form factors $F_{TA}$ and $F_{TV}$ are equal, 
$F_{TA}(E_{\mathrm{max}})=F_{TV}(E_{\mathrm{max}})$.  


\item In the region where $E\gg \Lambda_{\mathrm{QCD}}$, the form factors obey the LEET relation, 
which is  valid to $O(\Lambda_{\mathrm{QCD}}/m_b)$, $O(\Lambda_{\mathrm{QCD}}/E)$, 
and $O(\alpha_s)$ accuracy: 
\begin{eqnarray}
F_V\simeq F_A \simeq F_{TA}\simeq F_{TV}\propto  {f_B M_B}/{E}.
\end{eqnarray}

\item At large $q^2\simeq M_B^2$ (i.e., at small $E$), 
the following relation for the form factors should hold: 
\begin{eqnarray}
F_A\simeq F_{TA} \ll F_V \simeq F_{TV}.
\end{eqnarray}
We expect these relations to work with $10$--$15\%$ accuracy. 
\end{itemize}
Given these features, we now turn to the analysis of existing predictions for the $B\to\gamma$ form factors. 

\subsubsection{Form factors $F_A$ and $F_V$}

The form factors $F_A$ and $F_V$ for the $B_d\to\gamma$  transition 
have been calculated in 
Ref.~\cite{ms} within the dispersion approach of Ref.~\cite{m}.\footnote{
The form factors $F_{A,V}$ for the $B_u\to\gamma$ transition have been calculated
in Refs.~\cite{ali,eilam:etal} using light-cone sum rules. These results agree with the  
results from the dispersion approach \cite{ms}. 
It should be noted that the form factors $F_{A,V}$ for the $B_u\to\gamma$ 
transition have the opposite sign 
and are approximately twice as big as the corresponding 
form factors $F_{A,V}$ for the $B_d\to\gamma$ transition; 
see the discussion in Ref. \cite{ms}.} 
For large and intermediate values of $E$, these form factors
can be well parametrized by a particularly simple formula:
\begin{eqnarray}
\label{model}
F(E)=\beta\;\frac{f_B M_B}{\Delta+E}, 
\end{eqnarray}
with $f_B = 0.2$  GeV, $M_B=5.28$  GeV, and $\beta_V-\beta_A=O(1/m_b)$ in accord with 
LEET.
In order to use this formula for the form factors in the full
range of $E$, we should take into account that the form factors 
$F_V$ and $F_A$ have, respectively, poles 
at $q^2=M_{B^*}^2$ 
and $q^2=M_{B^{**}}^2$, so that  
$\Delta_V=M_{B^*}-M_{B}$ and $\Delta_A= M_{B^{**}}-M_{B}$.
Although the masses of the positive-parity states with higher spins $B^{**}$ 
are not known, we can use data on $D$ mesons to estimate the mass difference 
$\Delta_A\simeq 0.3$--$0.4$ GeV; here we take $\Delta_A=0.3$ GeV. 
The numerical parameters are listed in Table \ref{table:ffs}.
%
%

The maximal difference between the form factors at $E=0$ is at the level of 
$50\%$.  The difference between $F_A$ and $F_V$ is around $10\%$ for $E\ge 0.7$ GeV, 
indicating  
$\Lambda_{\mathrm{QCD}}/m_b$ corrections to the LEET relation between the 
form factors at the level of $5$--$10\%$.

\subsubsection{Form factors $F_{TA}$ and $F_{TV}$}
There are several calculations of the $B\to\gamma$ form factors $F_{TA}$ and 
$F_{TV}$ available in the literature \cite{aliev,geng}. 
Let us check whether the results for the form factors satisfy  the constraints 
derived above. 


(i) The light-cone sum rule calculation of Ref. \cite{aliev} predicts the form factors 
$F_{TV}\gg F_{TA}$ for large values of $E$, including $E_{\mathrm{max}}$, 
which points to a very strong violation of the LEET relations. 
More importantly, the {\it exact} relation in Eq.~(\ref{exact2}) between the form factors 
at $E_{\mathrm{max}}$ is also drastically violated.  
Thus, we conclude that the calculation of form factors performed in 
Ref.~\cite{aliev} cannot be trusted. 

(ii) The quark model calculation of Ref.~\cite{geng} satisfies the exact constraint 
in Eq.~(\ref{exact2}), with values of the form factors $F_{TA}=F_{TV}=0.115$
at $q^2=0$ (or, equivalently, $E_{\mathrm{max}}=M_B/2$) \cite{geng:private}. 
Taking into account the LEET relation (\ref{leet}), 
this value is in agreement with our results for the form factors $F_A=0.09$ and $F_V=0.105$ at $q^2=0$.  On the other hand, there are several features of the 
predicted  form factors that do not seem realistic.  First, as can be seen from
Fig.~1 of Ref.~\cite{geng},
the form factors $F_{TA}$ and $F_{TV}$ differ considerably, with 
$F_{TA}\gg F_{TV}$ for the values of $E\simeq 0.5$--$1$ GeV. This signals 
a very strong violation of the LEET condition, 
with corrections of the order of several hundred percent. 
Let us recall that the form factors may indeed be very different in the region 
$q^2\simeq M_B^2$, since $F_{TV}$ contains a pole at $q^2=M^2_{B^*}$ while
$F_{TA}$ does not. But then one would expect the relation $F_{TV}\gg F_{TA}$, 
opposite to the one obtained in Ref.~\cite{geng}. 
Second, the form factors $F_{TA}$ and $F_{TV}$ of Ref.~\cite{geng} vanish at 
$q^2=M_B^2$. 
Taking into account that the form factor $F_{TV}$ contains a pole at $q^2=M^2_{B^*}$, 
it seems very unlikely that the form factor $F_{TV}$ vanishes 
at $q^2=M_B^2$. Therefore, the predictions  of 
Ref.~\cite{geng} for  values of $q^2\ge 10$ GeV$^2$ cannot be considered very reliable. 

To sum up: There are no fully convincing results for the form factors 
$F_{TA}$ and $F_{TV}$; thus, for the analysis of the FB asymmetry, we prefer to 
rely upon a simple model for the $B\to\gamma$ transition form factors which 
satisfies  explicitly all the constraints discussed above. 

\subsection{Model for the $\bm{B\to \gamma}$ form factors} 
\begin{figure}
\begin{center}
\begin{tabular}{c}
\includegraphics[width=2.8in]{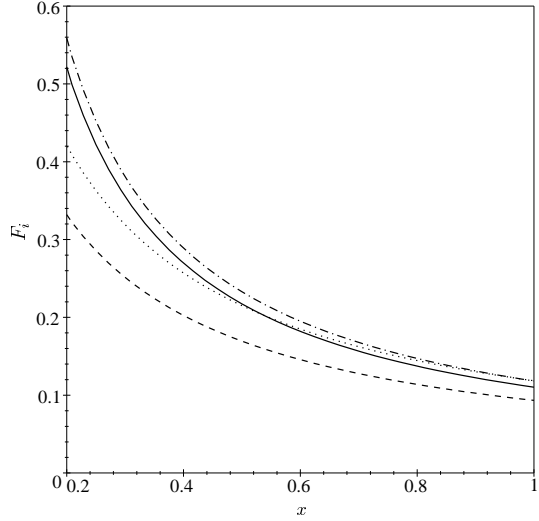}
\end{tabular}
\caption{\label{myffs}The predicted $x$ dependence of the $B_d\to\gamma$ 
form factors $F_{V}$ (solid curve), $F_{A}$ (dashed curve), $F_{TA}$ (dotted curve), 
and $F_{TV}$ (dash-dotted curve) according to our model, as described in the text ($x\equiv 2E/M_B$).}
\end{center}
\end{figure}

We assume the ansatz in Eq.~(\ref{model}) to be valid for all $B_{d,s}\to \gamma$ form factors with 
their own constants. 
Together with the condition in Eq.~(\ref{exact1}), this 
leads to the following relation between the parameters of 
$F_{TA}$ and $F_{TV}$:
\begin{eqnarray}
\frac{\beta_{TV}}{\Delta_{TV}+M_B/2}=\frac{\beta_{TA}}{\Delta_{TA}+M_B/2}, 
\end{eqnarray}
such that $\beta_{TV}-\beta_{TA}=O(1/m_b)$. 
Furthermore, according to our arguments mentioned above, we set 
\begin{eqnarray}
\Delta_{TV}=\Delta_V,\qquad \Delta_{TA}=\Delta_A. 
\end{eqnarray}
The remaining parameter to be fixed is the constant $\beta_{TV}$, for which we write 
\begin{eqnarray}
\beta_{TV}=(1+\delta)\beta_{V}, 
\end{eqnarray} 
and choose $\delta = 0.1$ according to the result of Ref.~\cite{geng}.  
This completes our simple model for the form factors which are 
consistent  with the exact relations 
at $E_{\rm max}$, LEET at large $E$, and heavy quark symmetry at small $E$. 
Table \ref{table:ffs} contains the various numerical parameters for the 
$B_d\to \gamma$ form factors, and Fig.~\ref{myffs} shows the 
form factors in our model as a function of the scaled photon energy, $x\equiv 2E/M_B$.  
%
%
\begin{table}[h]
\begin{tabular}{lcccc}
\hline\hline
Parameter                &   $F_V$   &  $F_{TV}$   &  $F_A$   &  $F_{TA}$ \\ \hline
$\beta$ (GeV${}^{-1}$)    &   0.28    &   0.30      &   0.26   &   0.33     \\
$\Delta$ (GeV)            &   0.04    &   0.04      &   0.30   &   0.30     \\
\hline\hline
\end{tabular}
\caption{Parameters of the $B_d\to\gamma$ form factors, as defined in Eq.~(\ref{model}).\label{table:ffs}}
\end{table}
%
%

For the $B_s\to\gamma$ transition, we do not know the precise values of the 
form factors, but we shall assume that the LEET-violating effects in the 
$B_s\to\gamma$ form factors have  the same
structure as those in the $B_d\to\gamma$ transition. With this assumption, the 
form factors as given in Table \ref{table:ffs} are sufficient
for the analysis of the FB asymmetry in the $B_s\to\gamma l^+l^-$ decay presented 
in the next section. 

\newpage
\section{Forward-backward  asymmetry in $\bm{B\to \gamma l^+ l^-}$}
We now assess the implications of our form factor model for the 
FB asymmetry of $\mu^-$ in the decay 
$\bar B_s\to \gamma \mu^+ \mu^-$.
Referring to Refs.~\cite{sehgal,aliev,eilam:etal,geng}, 
the radiative dilepton decay receives various contributions. 
The main contribution in the case of light leptons comes from the so-called 
structure-dependent (SD) part, where the photon 
is emitted from the external quark line.
Contributions coming from photons attached to charged internal lines 
are suppressed by $m_b^2/M_W^2$ \cite{eilam:etal}. 
The bremsstrahlung contribution 
due to emission of the photon from the external leptons is 
suppressed by the mass of the light leptons $l=e, \mu$ 
and affects the photon energy spectrum only in the low $E$ region \cite{sehgal,geng}.

Neglecting the bremsstrahlung contributions, the decay is then governed by the 
effective Hamiltonian describing the 
$b\to s l^+l^-$ decay, together with the form factors parametrizing the 
$B\to \gamma$ transition, as discussed in the preceding section.  
Using the effective Hamiltonian for $b\to s l^+ l^-$ in the SM \cite{heff:SM}, 
the matrix element is ($m_s=0$)  
\begin{eqnarray}
{\mathcal M}_{\mathrm{SD}}&=& 
\frac{G_F \alpha}{\sqrt{2} \pi}V_{tb}^{}V_{ts}^{\ast}
\Bigg[
\left(c_9^{\mathrm{eff}} \bar{l}\gamma^{\mu}l+c_{10} \bar{l}\gamma^{\mu}\gamma_5 l\right)
\langle {\gamma(k)}|{\bar{s}\gamma_{\mu}P_L b}|{\bar{B}_s (p)}\rangle
\nonumber\\
&&\hspace{5em}\mbox{}-\frac{2 c_7^{\mathrm{eff}}m_b}{q^2}
\langle{\gamma(k)}|{\bar{s}i \sigma_{\mu\nu}q^{\nu}P_R b}|{\bar{B}_s (p)}\rangle \bar{l}\gamma^{\mu}l\Bigg],
\end{eqnarray}
where $q=p-k$ and $P_{L,R}= (1\mp \gamma_5)/2$. Within the SM, the  Wilson coefficients, 
including next-to-leading-order corrections \cite{heff:SM}, 
have numerical values ($m_t=166\ {\mathrm{GeV}}$)
\begin{eqnarray}\label{wcs}
c_7^{\mathrm{eff}}= -0.330, \quad c_9^{\mathrm{eff}}= c_9 + Y(q^2), 
\quad c_9=4.182, \quad c_{10}= -4.234,
\end{eqnarray}
where the function $Y$ denotes contributions from the one-loop matrix elements of 
four-quark operators (see Ref.~\cite{heff:SM} for details), and has absorptive parts 
for $q^2> 4m_c^2$. 
In addition to the short-distance contributions,  there are also $c\bar{c}$ 
resonant intermediate states such as  $J/\psi, \psi'$, etc., which we will take 
into account by utilizing $e^+e^-$ annihilation data, as described in 
Ref.~\cite{res:exp}. 

Defining the angle $\theta$ between the three-momentum 
vectors of ${\mu^-}$ and the photon in the dilepton centre-of-mass system, 
and recalling the scaled energy variable $x \equiv 2E/M_{B_s}$
in the $B_s$ rest frame, we obtain the  differential decay rate 
($\hat{m}_i\equiv m_i/M_{B_s}$)
\begin{eqnarray}\label{diff}
\frac{d\Gamma(\bar B_s\to \gamma \mu^+ \mu^-)}{dx\, d \!\cos\theta} 
= \frac{G_F^2\alpha^3 M_{B_s}^5}{2^{11} \pi^4}|V_{tb}^{}V_{ts}^*|^2
\,x^3\Bigg(1-\frac{4\hat{m}_\mu^2}{1-x}\Bigg)^{1/2} [B_0(x) + B_1(x)\cos\theta + B_2(x) \cos^2\theta].
\end{eqnarray}
Here, we have summed over the spins of the particles in the final state, and have 
introduced the  auxiliary functions
\begin{eqnarray}
B_0 &=& (1-x + 4 \hat{m}_\mu^2)(F_1+F_2)- 8  \hat{m}_\mu^2 
|c_{10}|^2 (F_V^2+ F_A^2),\nonumber \\
B_1 &=& 8\Bigg(1-\frac{4\hat{m}_\mu^2}{1-x}\Bigg)^{1/2} 
\mathrm{Re\,}\{c_{10}[
c_9^{\mathrm{eff}*}(1-x) F_V F_A + 
c_7^{\mathrm{eff}}\hat{m}_b(F_V F_{TA} + F_A F_{TV})]\},   \nonumber\\
B_2&=& (1-x- 4\hat{m}_\mu^2)(F_1+F_2),
\end{eqnarray}
with the form factors defined in Eq.~(\ref{real}), and 
\begin{eqnarray}
F_1&=& (|c_9^{\mathrm{eff}}|^2 + |c_{10}|^2) F_V^2 + 
\frac{4|c_7^{\mathrm{eff}}|^2\hat{m}_b^2}{(1-x)^2}F_{TV}^2
 + \frac{4 \mathrm{Re\,}(c_7^{\mathrm{eff}} c_9^{\mathrm{eff}*})\hat{m}_b}{1-x} 
F_V F_{TV},\nonumber\\
F_2&=& (|c_9^{\mathrm{eff}}|^2 + |c_{10}|^2) F_A^2 + 
\frac{4|c_7^{\mathrm{eff}}|^2\hat{m}_b^2}{(1-x)^2}F_{TA}^2
+ \frac{4 \mathrm{Re\,}(c_7^{\mathrm{eff}} 
c_9^{\mathrm{eff}*})\hat{m}_b}{1-x} F_A F_{TA}.
\end{eqnarray}
Recall that the Wilson coefficient $c_9^{\mathrm{eff}}$ [Eq.~(\ref{wcs})] 
depends on $x$ via $q^2=M_{B_s}^2(1-x)$.

The term odd in $\cos\theta$ in Eq.~(\ref{diff}) produces a 
FB asymmetry, defined as\footnote{Note that the FB asymmetry is equivalent to the 
asymmetry in the $l^+$ and $l^-$ energy spectra discussed in Ref.~\cite{sehgal}. Our results in Eqs.~(\ref{diff}) and 
(\ref{FB:res}) agree with those given in that paper.}
\begin{eqnarray}\label{FB}  
A_{\text{FB}}(x)=\frac{\displaystyle\int_0^1 d\!\cos\theta
\frac{d\Gamma}{d x\, d\!\cos\theta}-\int_{-1}^0 d\!\cos\theta
\frac{d\Gamma}{d x\, d\!\cos\theta}}{\displaystyle\int_0^1 d\!\cos\theta
\frac{d\Gamma}{d x\, d\!\cos\theta}+\int_{-1}^0 d\!\cos\theta
\frac{d\Gamma}{d x\, d\!\cos\theta}},
\end{eqnarray} 
which is given by
\begin{eqnarray}\label{FB:res}
A_{\text{FB}}(x)= 3 \Bigg(1-\frac{4\hat{m}_{\mu}^2}{1-x}\Bigg)^{1/2}
\, \frac{\mathrm{Re\,}\{c_{10}[  c_9^{\mathrm{eff}*}(1-x) F_V F_A + 
c_7^{\mathrm{eff}}\hat{m}_b ( F_VF_{TA} + F_A F_{TV})]\}}{[(F_1 + F_2)(1-x + 2 \hat{m}_\mu^2) -6 \hat{m}_\mu^2 |c_{10}|^2(F_V^2+ F_A^2)]}.
\end{eqnarray}
Note that there are also non-factorizable 
radiative corrections to the asymmetry which are not contained in the transition 
form factors.\footnote{In the case of the $B\to K^*l^+l^-$ decay  
these corrections were analysed in \cite{beneke}.}

\begin{figure}
\begin{center}
\begin{tabular}{lr}
\includegraphics[width=2.8in]{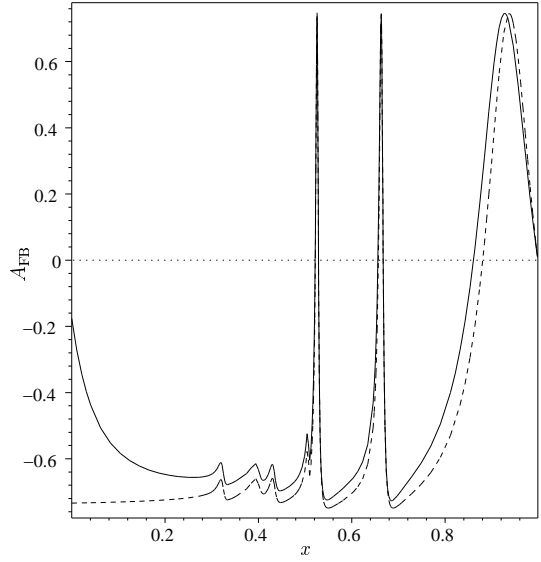} 
\end{tabular}
\caption{SM prediction for the 
FB asymmetry of $\mu^-$ in the decay $\bar{B}_s\to \gamma \mu^+\mu^-$
as a function of $x\equiv 2E/M_{B_s}$, using the form factors given in 
Eq.~(\ref{model}) (solid curve) and including the effects of $c\bar{c}$ resonances. 
For comparison, we also show the distribution obtained by utilizing the 
leading-order LEET form factor relation in Eq.~(\ref{leet}) (dashed curve). \label{fb:distr}}
\end{center}
\end{figure}

We plot in Fig.~\ref{fb:distr} the FB asymmetry as a 
function of the scaled photon energy $x$, by using our model for 
the form factors, Eq.~(\ref{model}), and the universal form factors, Eq.~(\ref{leet}). 
In  the latter case, omitting the non-factorizable corrections the dependence on 
the form factors drops out completely, and so the asymmetry is fully determined by 
the Wilson coefficients. 
%
%
%
%
%

From Fig.~\ref{fb:distr} one infers an interesting feature of  
$A_{\text{FB}}(x)$ in the SM: namely, for a given photon energy 
$x=x_0$, and far from the 
$c \bar{c}$ resonances,  the FB asymmetry vanishes.
%
As can be seen from Fig.~\ref{fb:distr}, the $1/M_B$ and $1/E$ corrections to the form factors, 
which are taken into account by our form factor model, Eq.~(\ref{model}), shift 
the location of the zero by only a few percent, and do not change the 
qualitative picture of the asymmetry for $x\gtrsim 0.4$. 
Using the numerical values of the standard model Wilson coefficients given in Eq.~(\ref{wcs}), 
together with $m_b=4.4\ \mathrm{GeV}$, we obtain $x_0\simeq 0.85$--$0.88$ 
depending on the form factors used. Notice again that the location of zero is
further affected by non-factorizable radiative corrections \cite{beneke}. 

We would like to emphasize that the absence of the zero in the SM 
forward-backward asymmetry in the region $x\ge 0.7$ reported in \cite{asym} 
is due to using the form factors of Ref.~\cite{aliev}, inconsistent with the 
rigorous constraints discussed in the present paper.
In view of this, the various distributions and asymmetries 
calculated in Refs.~\cite{bllg:decay,asym} with the form factors of Ref.~\cite{aliev} 
should be revised. 

\section{Conclusions}
We have analysed the form factors that describe the $B\to\gamma$ transition,
and investigated their implications for  the FB  asymmetry 
of the muon in the decay $\bar{B}_s\to \gamma \mu^+\mu^-$, within the SM.
Our results are as follows. 
\begin{itemize}
\item 
We have derived an exact relation for the form factors $F_{TA}$ and $F_{TV}$ of 
the $B\to\gamma$ transition induced by 
tensor and pseudotensor currents at maximum photon energy: 
\begin{eqnarray}
\label{rigorous}
F_{TA}(E_{\mathrm{max}})=F_{TV}(E_{\mathrm{max}}). 
\end{eqnarray}
\item 
We have investigated the resonance 
structure of the form factors at $q^2\simeq M_B^2$ and 
found that singularities of $F_T$ and $F_{TV}$ are located much closer to the 
edge of the physical region (i.e., $q^2=M_B^2$) than those of the form factors 
$F_{TA}$ and $F_{A}$. Hence we expect $F_T$ and $F_{TV}$ to rise 
rapidly as $q^2\to M_B^2$ but $F_A$ and $F_{TA}$ to remain relatively 
flat, so that at $q^2\simeq M_B^2$ we have the relation
\begin{eqnarray}
F_{A}\simeq F_{TA}\ll F_{V}\simeq F_{TV}. 
\end{eqnarray}
This behavior indicates a strong violation of the Isgur-Wise relations 
between the form factors at large $q^2$. 
\item 
We have found a serious discrepancy between the just-mentioned
constraints and existing calculations of the 
form factors $F_{TA}$ and $F_{TV}$ from QCD sum rules 
\cite{aliev} and quark models \cite{geng}: 

\noindent
(i) the form factors of Ref.~\cite{aliev} 
violate both the exact constraint (\ref{rigorous}) 
and the relations expected from the large energy effective theory (\ref{leet});  

\noindent
(ii) the form factors of Ref.~\cite{geng} 
signal a very strong violation of the LEET relation 
[Eq.~(\ref{leet})]. Moreover, 
the vanishing of the form factors $F_{TA}$ and $F_{TV}$ at $q^2=M_B^2$ 
in \cite{geng} contradicts 
the resonance structure of these form factors. 
\item
We would like to stress that there is an important relation between the 
universal form factors describing the $B_u\to\gamma$ and $B_d\to\gamma$ transitions: 
\begin{eqnarray}
\zeta^{B_u\to\gamma}_\perp(E,M_B)=Q_u/Q_d\;\zeta^{B_d\to\gamma}_\perp(E,M_B),  
\end{eqnarray}
where $Q_{u,d}$ represent the charge of $u$ and $d$ quarks. As a consequence 
of this relation,  the form factors $F_{A,V,TA,TV}$ of the $B_u\to\gamma$ transition 
have opposite sign, and their moduli are approximately twice as big as the
corresponding form factors of the $B_d\to\gamma$ transition. 
Furthermore, it is worth emphasizing that this relation has not been 
properly taken into account in Ref.~\cite{aliev} when using  
the $B_u\to\gamma$ form factors of Ref.~\cite{eilam:etal} for the description of the 
$B_{d,s}\to\gamma l^+l^-$ decay.  

\item
By using the exact relation between the form factors $F_{TA}$ and $F_{TV}$ 
at $E_{\rm max}$, the resonance structure of the form factors in the region 
$q^2\simeq M_B^2$, and the LEET relations 
$F_{A}\simeq F_{V}\simeq F_{TA}\simeq F_{TV}$ valid for $E\gg \Lambda_{\mathrm{QCD}}$,
we proposed a simple parametrization for the form factors:  
\begin{eqnarray}
F_i(E)=\beta_i\frac{M_B f_B}{\Delta_i + E}, \qquad i=A,V,TA,TV. 
\end{eqnarray}
The numerical parameters (Table \ref{table:ffs}) have been fixed by 
utilizing  reliable data on the form factors at large and intermediate values of the 
photon energy. 

\item
We have applied our form factor model to the FB asymmetry 
of the muon in $\bar{B}_s\to\gamma \mu^+ \mu^-$ decay. Comparing the distribution 
of $A_{\mathrm{FB}}$ based on these form factors with the one obtained 
by using the leading-order LEET form factors shows that the behavior of 
the FB asymmetry remains essentially unchanged in the region $x=2E/M_{B_s}\gtrsim 0.4$. 

Our analysis confirms the result of Ref.~\cite{sehgal} 
that the shape of the FB asymmetry as well as the location of its zero are 
typical for the SM. 
We point out that the asymmetries and distributions reported in a number of 
recent publications \cite{bllg:decay,asym} should be revised as they are 
based on the form factors of Ref.~\cite{aliev} which are inconsistent with the 
rigorous constraints on the form factors. 
\end{itemize}

According to the above results we conclude that the FB asymmetry in the
$B\to \gamma l^+l^-$ decay, particularly its zero arising in the SM, can
be predicted with small theoretical uncertainties. This is similar to the
decay $B\to K^*  l^+l^-$, where a full next-to-leading-order calculation
(second reference in \cite{beneke}) shows that a measurement of the zero of the FB asymmetry
would allow a determination of $c_7^{\rm eff}/{\rm Re}(c_9^{\rm eff})$ at the 10\% level. (This
order of magnitude should also hold in the case of the $B\to \gamma l^+l^-$ decay.)

To sum up, the study of the decay $B_s\to \mu^+\mu^-\gamma$ at future
hadron collider experiments will provide complementary information on the
structure of the underlying effective Hamiltonian describing $b \to s l^+l^-$ 
transitions.

\begin{acknowledgments}
We are grateful to Berthold Stech for valuable discussions. 
F.K.~has been supported by the Deutsche Forschungsgemeinschaft (DFG) 
under contract Bu.706/1-1. D.M. would like to thank the Alexander von Humboldt-Stiftung 
for financial support. 
\end{acknowledgments} 


\end{document}